\documentclass[11pt,reqno]{amsart}
\usepackage{amssymb,graphicx}

\usepackage{amsfonts,amscd,amsthm,amsmath}

\usepackage{extarrows}

\usepackage{cite}

\numberwithin{equation}{section}

\begin{document}

\begin{center}

\textbf{\large{Polypseudologarithms revisited}}

\vskip 5mm\textbf{Djurdje Cvijovi\'{c}}

\vskip 2mm {\it Atomic Physics Laboratory, Vin\v{c}a Institute of
Nuclear Sciences \\
P.O. Box $522,$ $11001$ Belgrade$,$ Republic of Serbia}\\
\textbf{E-Mail: djurdje@vinca.rs}\\

\vskip 2mm
\begin{quotation} \textbf{Abstract.} Lee, in a series of papers, described a unified formulation of the statistical
thermodynamics of ideal quantum gases  in terms of the polylogarithm functions,
$\textup{Li}_{s} (z)$. It is aimed here to investigate the functions  $\textup{Li}_{s} (z),$ for
$s = 0, -1, -2, \ldots,$ which are, following Lee, referred to as the polypseudologarithms (or polypseudologs)
of order $n$. Various known results regarding polypseudologs, mainly obtained in widely differing  contexts and currently scattered throughout the literature, have been brought together along with many new results and insights
and they all have been proved in a simple and unified manner. In addition, a new  general explicit closed-form
formula for these functions involving the Carlitz--Scoville higher tangent numbers has been established.
\end{quotation}

\end{center}




\noindent
\section{Introduction}

Lee,  in a series of papers \cite{Lee00,Lee01,Lee02,Lee03,Lee04,Lee05}, not only has shown for the first time  that statistical thermodynamics of the Bose--Einstein and Fermi--Dirac ideal gases can be unified, but also has obtained a unified description of the main thermodynamical functions in terms of the polylogarithmic functions $\textup{Li}_{s} (z)$ (for their definitions see Sec. 2) and also examined details of certain physical phenomena revealed by  this approach.

The unification is based on the following expression for the reduced density of non-relativistic ideal gas of $N$ particles confined to a  $d$--dimensional box of $\textup{volume} \,V$
\begin{equation} \frac{\rho \lambda^d }{g} = \textup{sgn} (\zeta) \, \textup{Li}_{d/2} (\zeta),\qquad\zeta = \left\{\begin{array}{l}
 \;\;\, z\qquad\textup{if}\;\;\textup{Bose--Einstein}
 \\
 - z\qquad\textup{if}\;\;\textup{Fermi--Dirac},
 \\
 \end{array} \right.
\end{equation}
\noindent where  $\rho\equiv N/V = N/L^d$ is  the particle number density,  $\lambda = (2\pi \beta/m)^{1/2}$  (with $\hbar =1$) is the thermal wave--length, $\beta\equiv 1/k_B T$ and $g = 2s + 1 $, while $k_B,$  $T$, $m$ and $s$ denote the Boltzman constant, the temperature, the mass and the spin of the particles, respectively. Finally,  the fugacity $z$ is related to the chemical potential $\mu$ of the  system as $z\equiv e^{\beta \mu}$. The grand partition function then easily  follows from (1.1) $Q =\exp\left\{\textup{sgn}(\zeta)  (\lambda/L)^d\, \textup{Li}_{d/2+1}(\zeta)\right\},$ while the basic thermodynamic functions, such as the energy $U$, the entropy $S$, the pressure $P$ and also fluctuation quantities such as the number fluctuations $Y$, can be expressed in polylogarithms because of their relationship with the density $\rho$.

The important property of the unified formulation is that $U,$ $S,$ $P,$  $Y$ and the reduced density depend only on two variables, the fugacity and dimensionality,  and, more importantly, their values are given by the very same  function -- polylogarithms (of order $s$ and argument $z$), $\textup{Li}_{s} (z)$; the fugacity $z$ determines the argument, while  the dimensionality $d$ enters the polylogarithms through its order--an integral order if $d$ is even, and a half-integral order if $d$ is odd.  Moreover, by the principle of analytic continuation, the obtained results may be assumed valid for any $d$.

It is noteworthy that the polylogarithm formulation of statistical thermodynamics evidently reveals that the thermodynamic properties of ideal quantum gases are described by the structural properties of $\textup{Li}_{s} (z)$ and, therefore, there is a need to understand better these important but little known functions. In addition, it is not surprising that this formulation  is a quite natural framework for dealing with dimensionality effects in ideal quantum gases, such as anomalous physics in $d = 0$ \cite{Lee03} and a remarkable  thermodynamic  equivalence (discovered by Lee) between the Bose--Einstein and Fermi--Dirac gases in $d = 2$ \cite{Lee04, Patton}. There are suggestions that negative dimensionality, {\em i.e.}  $d < 0,$
might be  of some theoretical interest \cite{Lee05, Baram}.

Motivated by the work of Lee \cite{Lee00,Lee01,Lee02,Lee03,Lee04,Lee05} on the thermodynamics of quantum  gases in terms of polylogarithms and, in particular, by his paper on {\it polypseudologarithms}$,\,$ {\em i.e.} $\textup{Li}_{s} (z)$  for $s = 0, -1, -2, \ldots,$ etc.  \cite[Sec. 3]{Lee05}, we have begun a systematic study of $\textup{Li}_{s} (z)$ of non-positive integral order $s$. Various known results regarding these functions, mainly obtained in different contexts and currently scattered throughout the literature, have been brought together here along with many new results and insights and they all have been proved in a simple and unified manner. However, it is neither intended nor attempted to provide a detailed historical account of this topic, nor to attribute the results to the original authors. Much attention has been given to considering all possible ways of computing $\textup{Li}_{s} (z)$  for $s = 0, -1, -2, \ldots,$ and, in particular, to the problem raised by Lee \cite[Sec. 3]{Lee05}  concerning the existence of a
general closed-form expression for these functions. It has turned out that there are, in the mathematical literature,
 several such expressions involving either the Stirling numbers of the second kind or the  Eulerian numbers. In Sec. 4 we shall deduce a new explicit closed-form expression  involving the Carlitz-Scoville tangent numbers.

\section{Polypseudologs, their properties and generation }

The polylogarithm functions (polylogarithms, or polylogs for short) of  order $s$ and argument $z,$ $\textup{Li}_{s} (z),$ are defined by
\begin{equation*}\textup{Li}_{s} (z) = \frac{z}{\Gamma(z)} \int_{0}^{1} \left[\log(1/t)\right]^{s-1} \frac{dt}{1- zt},\qquad(\Im{(t)} = 0)
\end{equation*}
\noindent whenever this integral converges, {\em i.e.,} $\Re{(s)}>0,$ $\Re{(z)}<1,$ and elsewhere by analytic continuation. It is assumed that $\log(1/t)$ has its principal value and, evidently, there is branch cut from $z=1$ to $\infty$. In the important case where the parameter $s$ is an integer, $\textup{Li}_{s} (z)$ will be denoted by $\textup{Li}_{n} (z)$ (or $\textup{Li}_{-n} (z)$ when the parameter is negative).  $\textup{Li}_{n} (z),$ $n\in\mathbb{N}:=\{1, 2, \ldots,\},$ are the ({\em classical}) polylogarithms of order $n$ ({\em i.e.} the $n$th order polylogarithms). The special case $n = 1$ is the ordinary logarithm (or monologarithm) $\textup{Li}_{1} (z)= -\log(1-z),$ while the cases $n = 2, 3, 4,\ldots,$ are known, respectively, as dilogarithm, trilogarithm, quadrilogarithm, etc.

For more details and an extensive list of references in which polylogarithms appear in physical and mathematical problems we refer the reader to Maximon \cite{Maximon}. The $n$th order polylogarithms are thoroughly covered in Lewin's standard text \cite{Lewin}, while many formulae involving $\textup{Li}_{s} (z)$  can be found in  Erd\'elyi {\em et al.} \cite[pp. 30--31]{Erdelyi} and Prudnikov {\em et al.} \cite[pp. 762--763]{Prudnikov}. For some new results and applications see \cite{Cvijovic1, Lee06,Cvijovic2}.

Following Lee \cite{Lee05}, the functions $\textup{Li}_{-n} (z),$ $n\in\mathbb{N}_{0},$ are here referred to as the polypseudologarithms (or polypseudologs)  of order $n$ and we first summarize their main properties and later thoroughly consider different ways of their computation. Polypseudologs  exhibit several very important  properties which are common to all polylogarithms.   ({\em i}) The recurrence relation for polylogs  can be rewritten as follows
\begin{equation}z\,\frac {\textup{d}}{\textup{d} z} \,\textup{Li}_{-n} (z)= \textup{Li}_{-(n+1)} (z)\qquad(n\in\mathbb{N}_{0}).
\end{equation}
\noindent ({\em ii}) Polypseudologs  satisfy the following particularly simple inversion relation
\begin{equation}\textup{Li}_{-n} \left(\frac{1}{z}\right)= (-1)^{n+1} \,\textup{Li}_{-n} (z)\qquad(n\in\mathbb{N}).
\end{equation}
\noindent ({\em iii}) Duplication relation (or quadratic transformation)
\begin{equation}\textup{Li}_{-n} (z)+ \textup{Li}_{-n} (-z) =2^{1+n} \textup{Li}_{-n} (z^2)
\end{equation}
\noindent is satisfied.  ({\em iv}) If $|z|<1$,  polypseudologs have the series expansions  given by $\sum_{k\,=1}^{\infty} k^{n} z^n$. In addition, $\textup{Li}_{-n} (z)$ have some additional, to these functions, specific properties. ({\em v}) $\textup{Li}_{0} (z),$ $\textup{Li}_{-1} (z),$ $\textup{Li}_{-2} (z),$ $\ldots ,$ are rational functions. ({\em vi}) They have a pole of order $n+1$ at $z=1$. ({\em vii}) The values of polypseudologs at $z=-1$ are related to the values of the Riemann zeta function $\zeta(s)$ at negative integers and are expressible in terms of the Bernoulli numbers $B_n$ \begin{equation}\textup{Li}_{-n} (-1) = (2^{1+n} -1)\, \zeta(-n) = (1-2^{1+n}) \,\frac{B_{n+1}}{n+1}\qquad(n\in\mathbb{N}).
\end{equation}
\noindent Hence, $\textup{Li}_{-2 n} (-1) = 0$  and $\textup{Li}_{1-2 n} (-1) = (1-2^{2 n}) B_{2 n}/(2 n)$ since  $B_{2 n+1}= 0.$  ({\em viii})  $\textup{Li}_{-n} (z)$  are, for all orders, factorable by $z$ and also by $(z+1)$ if $n$ is an even number.

Lee \cite[Sec. III]{Lee05}, by applying the recurrence relation to $\textup{Li}_{1} (z)= -\log(1-z)$ and repeating it over and again, obtained the $\textup{Li}_{-n} (z)$ to order $n = 8$:
\begin{align*}
&\textup{Li}_{0} (z) = z/(1-z),\hskip110mm
\\[-1mm]
&\textup{Li}_{-1} (z) = z/(1-z)^2
\\[-1mm]
&\textup{Li}_{-2} (z) = z (1+z)/(1-z)^3,
\\[-1mm]
&\textup{Li}_{-3} (z)= z (1+ 4 z +z^2)/(1-z)^4,
\\[-1mm]
&\textup{Li}_{-4} (z)= z (1+z) (1+ 10 z + z^2)/(1-z)^5,
\\
&\textup{Li}_{-5} (z)= z (1+ 26 z + 66 z^2 +26 z^3+z^4)/(1-z)^6,
\\
&\textup{Li}_{-6} (z)= z (1+z)(1+ 56 z + 246 z^2 + 56 z^3 + z^4)/(1-z)^7,
\\
& \textup{Li}_{-7} (z)= z (1+ 120 z + 1191 z^2 + 2416 z^3 + 1191 z^4 +120 z^5 + z^6)/(1-z)^8,
\\
&\textup{Li}_{-8} (z)= z (1+z) (1+ 246 z + 4047 z^2 + 11572 z^3 + 4047 z^4 + 246 z^5 + z^6)/(1-z)^9.
\end{align*}
\noindent Clearly, in this way one can find the polylog of any desired lower order. In this section we consider a half-dozen additional ways for obtaining polypseudologs and majority of them are the closed-form formulae.

The second method  to generate $\textup{Li}_{-n} (z)$ will next be described.  Note that $\textup{Li}_{-1} (z) = \left(z \frac{\textup{d}}{\textup{d} z}\right) \sum\nolimits_{k\,=0}^{\infty} z^k$, where we first differentiate and then multiply by $z$, and,  if we apply the operator  $z \frac{\textup{d}}{\textup{d} z}$ $n$ times, which we denote by the symbol $\left(z \frac{\textup{d}}{\textup{d} z}\right)^n$, to the summation formula $\sum\nolimits_{k\,=0}^{\infty} z^k = (1-z)^{-1}$, we have \cite[p. 364, Eq. (2)]{Knopf}
\begin{equation}\textup{Li}_{-n} (z)= \left(z \, \frac {\textup{d}}{\textup{d} z}\right)^n \frac{1}{1 - z}\qquad(n\in\mathbb{N}).
\end{equation}
\noindent Moreover, it will be shown below that this relation leads to a closed-form formula involving the Stirling numbers of the second kind.

An alternative  way of generating $\textup{Li}_{-n } (z)$ for any $n$ would be to make use of the generating function method, {\it i.e.} to generate $\{\textup{Li}_{-n } (z)\}_{n\,=1}^{\infty}$  from a single function of two variables $G(z,t)$ by repeated differentiation of that function. It is fortunate that there are several such functions (of which (2.6a) and (2.6b) could be found  in the literature; see \cite[p. 987]{Zeitlin} and \cite[p. 152, Eq. (19)]{Truesdell})
\begin{align*} &\sum_{n\,=1}^{\infty}\frac{t^n}{n!}\, \textup{Li}_{-n} (z)=\frac{1}{1-z\,e^t},\tag{2.6a}
\\
&\sum_{n\,= 0}^{\infty} \frac{t^n}{n!} \,(-1)^n  \textup{Li}_{-n}(z)=\frac{z}{e^t -z},\tag{2.6b}
\\
&\sum_{n\,= 2}^{\infty} \frac{t^n}{n!} \,(-1)^{n-1}  \textup{Li}_{-(n-1)}(z)=\log(e^t -z),\tag{2.6c}\hskip50mm
\\
&\sum_{n\,= 1}^{\infty}\frac{t^n}{n!}\, \textup{Li}_{-n} (z)= \frac{1}{2}\,\frac{1+z\,e^t}{1-z\,e^t},\tag{2.6d}
\end{align*}
\noindent so that, for $n\in\mathbb{N}$, we have
\begin{align*}\textup{Li}_{-n} (z) & \xlongequal{\textup{a}} \left. \frac {\textup{d}^n}{\textup{d} t^n} \left(\frac{1}{1- z \,e^t}\right)\right|_{t =0}^{} \xlongequal{\textup{b}} \left.(-1)^n z \,\frac {\textup{d}^n}{\textup{d} t^n}\left( \frac{1}{e^t - z }\right)\right|_{t =0}^{}\nonumber
\\\tag{2.7}
& \xlongequal{\textup{c}} \left.(-1)^n  \frac {\textup{d}^{n+1}}{\textup{d} t^{n+1}}\,\log(e^t-z)\right|_{t =0}^{}\xlongequal{\textup{d}} \frac{1}{2}\,\left. \frac {\textup{d}^n}{\textup{d} t^n} \left(\frac{1+z \,e^t} {1- z \,e^t}\right)\right|_{t =0}^{}.\nonumber
\end{align*}

Observe that it is in fact  enough only to demonstrate  that (2.6a) is the generating function for $\textup{Li}_{-n}(z);$ the function in (2.6b) follows from (2.6a) (as well as {\it vice versa}) by appealing to the inversion property in (2.2), while,  the functions given by (2.6b) and (2.6c) could be transformed into each other by the recurrence relation (2.1). It is interesting that both  functions $1/(1- z e^t)$ and $(1/2) (1+ z e^t)/(1 - z e^t)$  have the same first derivative $e^t/(1- z e^t)^2$, thus we have that (2.6a) and (2.6d) are equivalent. Regarding (2.6a), let $z e^t <1,$ where $\left|z\right|<1,$ then, since  $\sum\nolimits_{k\,= 0}^{\infty }(z e^t)^k = 1/(1 - z e^t)$, it is obvious that $n$th derivative of $1/(1- z e^t)$ with respect to $t$ evaluated at $t = 0$ gives $\textup{Li}_{-n}(z)$.

Another way to compute $\textup{Li}_{-n}(z)$ stems from the fact that all the derivatives indicated in (2.7) are readily expressible analytically in a closed form. Indeed, if we use the known result (\cite[p. 152]{Truesdell} and \cite[p. 32]{Jordan})
\begin{equation}
\frac {\textup{d}^n \, \Phi(e^t)}{\textup{d} t^n}\Big|_{t =0}^{} = \sum_{k\,= 1}^n S (n,k)\, \frac {\textup{d}^k \, \Phi(t)}{\textup{d} t^k }\Big|_{t = 1}^{}\qquad(n\in\mathbb{N}),\tag{2.8}
\end{equation}
\noindent where $S(n,k)$ stands for the Stirling numbers of the second kind defined by means \cite[p. 204, Eq. (1b)]{Comtet}
\begin{equation}S(n,k)= \frac{1}{k!} \sum_{\ell\,= 0}^{k} (-1)^{k-\ell} \binom{k}{\ell}\ell^n = \frac{1}{k!} \sum_{\ell\,= 0}^{k} (-1)^{\ell} \binom{k}{\ell}(k - \ell)^n,\tag{2.9}
\end{equation}
\noindent (also see (2.12) below) and $\Phi(t)$ is any infinitely differentiable function, then, starting  from the corresponding expressions in (2.7), we arrive at the following explicit formulae valid for $n\in\mathbb{N}$ (for (2.10b), see \cite[p. 152]{Truesdell})
\begin{align} & \textup{Li}_{-n } (z)=\sum_{k\,= 1}^n \frac{k! \,S(n,k)\,z^k}{(1-z)^{k+1}},\tag{2.10a}
\\
& \textup{Li}_{-n } (z)=\sum_{k\,= 1}^n \frac{ (-1)^{n+k} k! \, S(n,k)\,z}{(1-z)^{k+1}},\tag{2.10b}
\\
& \textup{Li}_{-n } (z)=\sum_{k\,=0}^n \frac{(-1)^{n+k} k! \,S(n+1,k+1)}{(1-z)^{k+1}}.\hskip 45mm\tag{2.10c}
\end{align}
\noindent We remark that (2.10a) is obtained by (2.7a) as well as by (2.7d), and, furthermore, this formula also follows from (2.5) upon using  \cite[p. 364, Eq. (3)]{Knopf}
\begin{equation}
\left(z\,\frac {\textup{d}}{\textup{dz}}\right)^{n} \Psi(z)= \sum_{k\,= 1}^n z^k \, S(n,k)\, \frac {\textup{d}^k \, \Psi(t)}{\textup{d}z^k }\qquad(n\in\mathbb{N}),\tag{2.11}
\end{equation}
\noindent $\Psi(z)$ being any infinitely differentiable function. It should be noted that it is easy to prove the formulae (2.8) and (2.11) by induction on $n.$ For $n =1,$ they are obviously true and all that is needed for the induction step is the  recurrence relation for $S(n,k)$ \cite[p. 208, Eq. (3a)]{Comtet}
\begin{align}
&S(n,k)= k \,S(n - 1,k)+ S(n -1, k-1),\nonumber
\\
&S(n,0) = S(0,k)= 0 \quad\textup{and}\quad  S(0,0)=1 \hskip 25mm(k,n\in\mathbb{N}).\tag{2.12}
\end{align}

Further, in view of the value of $S(n,k)$ explicitly given by (2.9), it is straightforward, starting from  (2.10), to obtain another group of the closed-form formulae, this time involving double finite sums (for (2.13c), see \cite[p. 988, Eq. (7)]{Zeitlin})
\begin{align} & \textup{Li}_{-n } (z) = \sum_{k\,= 1}^n \sum_{\ell\,= 0}^k (-1)^{k- \ell} \binom{k}{\ell} \frac {\ell^n \,z^k}{(1-z)^{k+1}},\tag{2.13a}
\\
 & \textup{Li}_{- n } (z)= \sum_{k\,= 1}^n \sum_{\ell\,= 0}^k (-1)^{n -\ell} \binom{k}{\ell} \frac {\ell^n \,z}{(1-z)^{k+1}},\tag{2.13b}
\\
 & \textup{Li}_{-n } (z)= \sum_{k\,= 0}^n \sum_{\ell\,= 0}^{k+1} \frac{(-1)^{n +1 -\ell}}{k+1} \binom{k +1}{\ell}\frac {\ell^{n+1}}{(1-z)^{k+1}}.\hskip 25mm\tag{2.13c}
\end{align}

Next, it is not difficult to show that, in terms of the Eulerian polynomials and numbers, $A_n(z)$ and  $A(n,k)$,  we have \cite[p. 245, Eq. (5n)]{Comtet}
\begin{equation}\textup{Li}_{-n } (z) = \frac{A_n(z)}{(1-z)^{n+1}} = \frac{1}{(1-z)^{n+1}}\,\sum\limits_{k\,=0}^{n} A(n,k)\, z^k.\tag{2.14}
\end{equation}

\noindent The {\it Eulerian polynomials}, $A_n(x),$  are usually defined by the generating function \cite[p. 244, Eq. (5i)]{Comtet}
\begin{equation}\sum_{n\,=0}^{\infty} A_n(x)\,\frac{t^n}{n!} = \frac{1-x}{1- x\, e^{t (1-x)}}\tag{2.15}
\end{equation}
\noindent and they are polynomials in $x$ of degree $n,$ explicitly given by
\begin{equation}A_n(x)= \sum_{k\,= 0}^{n} A(n,k)\, x^{k},\tag{2.16}
\end{equation}
\noindent where the coefficients $A(n,k)$ are the {\it Eulerian numbers}  \cite[p. 242, Eq. (5f)]{Comtet}
\begin{equation}A(n,k)= \sum_{\ell\,=0}^{k} (-1)^{\ell}\binom{n + 1}{\ell} (k-\ell)^n.\tag{2.17}
\end{equation}
\noindent Observe that  the Eulerian polynomials and numbers are not to be confused with the Euler polynomials $E_n(x)$ and numbers $E_n$.

To prove (2.14) we may proceed as follows: upon replacing  $t$ by $t/(1-x)$ in (2.15) and dividing both sides of the obtained expression by $(1-x)$, the right-hand side of (2.15) then becomes $1/(1 - x e^t)$ which is the same as the right-hand side of (2.6a), and hence (2.14).

However, there is one more formula: by (2.14) and (2.17) we get (\cite[p. 325]{Stalley}, \cite{Klamkin} and \cite[p. 987, Eq. (3)]{Zeitlin})
\begin{equation}\textup{Li}_{-n } (z) = \frac{1}{(1-z)^{n+1}}\,\sum\limits_{k\,=0}^{n} \sum\limits_{\ell\,=0}^{k} (-1)^{\ell}\binom{n + 1}{\ell} (k-\ell)^n\, z^k.\tag{2.18}
\end{equation}

\vskip 5mm
\section{New explicit closed-form formula for  polypseudologs}

In this section we shall deduce the following new explicit closed-form formula for polypseudologs $\textup{Li}_{-n } (z),$ $n\in\mathbb{N},$
\begin{equation}\textup{Li}_{-n } (z) = \frac{1}{2^{n+1}} \left[(-1)^{\lfloor n/2- 1\rfloor}\, T(n,1) + \sum_{k\,= 1}^{n+1} \frac{(-1)^{\lfloor(n-k)/2-1\rfloor}}{k}T(n+1,k)\left(\frac{1 + z}{1 - z}\right)^k\right],
\end{equation}
\noindent where $T(n,k)$ stands for the tangent numbers (of order $k$) or the higher order tangent numbers  defined by \cite[p. 428]{Carlitz}
\begin{equation}
\tan^{k}(t)=\sum_{n = k}^{\infty}
T(n,k)\,\frac{t^n}{n!}\qquad(k\in\mathbb{N}).
\end{equation}

Our derivation of (3.1) goes through  two steps. The first step is to show that $\textup{Li}_{-n } (z)$ could be expressed in terms of  the {\em derivative polynomials for tangent}, $P_n(x),$ which are  given by means of the exponential generating function  \cite{Hoffman}
\begin{equation}
P(x,t):=\frac{x+\tan (t)}{1-x \tan (t)}= \sum_{n=0}^{\infty}
P_n(x)\,\frac{t^n}{n!}
\end{equation}
\noindent or,  equivalently,  by the recurrence relation \cite{Hoffman}
\begin{equation}
P_{0}(x) = x,\qquad P_{n}(x)=(1+x^2)\, P_{n-1}^{'}(x)\qquad(n\in\mathbb{N}).\tag{3.3*}
\end{equation}
\noindent The second step is to make use of a new formula for $P_n(x)$
\begin{equation}
P_n(x)= T(n,1) +\sum_{k\,=1}^{n+1} \frac{1}{k}\,T(n+1,k)\,x^k\qquad(n\in\mathbb{N}),
\end{equation}
\noindent which shall be  proved separately at the end of the section.

We are now ready to proceed with the derivation of (3.1). First, consider the generating function for the derivative polynomials $P_{n}(x)$ in (3.3). Upon multiplying both sides by $\imath/2,$ $\imath:=\sqrt{-1},$ replacing $x$ by $-\imath (1+z)/(1-z),$ replacing $t$ by $\imath t/2$ and noticing that $\tan(\imath t/2) = - \imath \,(1- e^t)/(1+e^t),$ we obtain, after simple algebra, the following
\begin{equation}\frac{1}{2} \cdot\frac{1 + z e^t}{1 - z e^t} = \sum_{n\,= 1}^{\infty} \left(\frac{\imath}{2}\right)^{n+1}\,P_n\left(\frac{1}{\imath} \cdot \frac{ 1+z}{1- z}\right)\frac{t^n}{n!}.
\end{equation}
Now, if we compare this with the generating function for $\textup{Li}_{-n } (z)$ given by (2.6d), then, clearly, we have
\begin{equation}\textup{Li}_{-n } (z) = \left(\frac{\imath}{2}\right)^{n+1}\,P_n\left(\frac{1}{\imath} \cdot  \frac{ 1+z}{1- z}\right)\qquad(n\in\mathbb{N}).
\end{equation}

Secondly, after combining (3.6) and (3.4), we get a formula similar to the required formula (3.1) and what remains is to evaluate the powers $\imath^{n + 1}$ and $\imath^{n - k + 1}$ which respectively appear  as multiplication factors before $T(n,1)$ and $T(n+1,k)$. At this point it is helpful to examine the tangent numbers in more detail (see Table 1 and Eq. (3.2)) and notice that $T(n,k)\neq 0$ only when  $1\leq k\leq n $ and either both $n$ and $k$ are even or both $n$ and $k$ are odd. In other words, $T(2 m, 2 r + 1)= 0$ and $T(2 m + 1,2 r) = 0$ $(m, r \in \mathbb{N}_{0})$.

\noindent  \begin{table}[ht]
\caption{Tangent numbers  $T(n,k)$} 
\centering 
\small
\begin{tabular}{c| c c c c c c c c c} 
$n\backslash k $ & 1 & 2 & 3 & 4 & 5 & 6 & 7 & 8 & 9 \\ [.5ex] 
\hline  
1 & 1 \\ 
2 & 0 & 2 \\
3 & 2 & 0 & 6 \\
4 & 0 & 16 & 0 & 24 \\
5 & 16 & 0 & 120 & 0 & 120 \\
6 & 0 & 272 & 0 & 960 & 0 & 720 \\
7 & 272 & 0 & 3\,696 & 0 & 8\,400 & 0 & 5\,040 \\
8 & 0 & 7\,936 & 0 & 48\,384 & 0 & 80\,640 & 0 & 40\,320 \\
9 & 7\,396 & 0 & 168\,960 & 0 & 645\,120 & 0 & 846\,720 & 0 & 362\,880 \\[1ex] 
\hline 
\end{tabular}
\label{table:nonlin} 
\end{table}

This means that the terms involving $T(n,1)$ when $n$ is odd and $T(n+1,k)$ when $n-k$ is odd are the only surviving terms in the obtained formula. Hence, we  only need to evaluate the powers $\imath^{n+1}$ when $n$ is odd and $\imath^{n - k +1}$ when $n-k$ is odd. This is how we have found the negative powers, so the proof of (3.1) is now complete.

In order to prove  formula (3.4) note that the generating function of  $P_n(x)$ in (3.3) can be rewritten as
\begin{equation*}
P(x,t) = \left[ x + \tan (t)\right]\,\sum_{k\, = 0}^{\infty}
\left[ x\,\tan(t)\right]^k = x+ \big(1+x^2\big)\,\sum_{k\,=1}^{\infty}
x^{k-1} \tan^k(t)
\end{equation*}
\noindent which, by making use of the definition of  $T(n,k)$ in  (3.2) and the following elementary double series identities \cite[p. 57, Eq. (2)]{Rainville}
\[\sum_{n\,=1}^{\infty}\sum_{k\,=1}^{n}a(k,n)=\sum_{n\,=1}^{\infty}\sum_{k\,=1}^{\infty}a(k,n+k)=
\sum_{n\,=1}^{\infty}\sum_{k\,=n}^{\infty}a(k,n),
\]
\noindent  becomes
\begin{align}
P(x,t) &= x + \big(1+x^2\big)\,\sum_{k=1}^{\infty}
x^{k-1} \,\sum_{n=k}^{\infty}
T(n,k)\,\frac{t^n}{n!}\nonumber
\\
&= x + \sum_{n=1}^{\infty}\big(1+x^2\big)\,\left(\sum_{k=1}^{n}
T(n,k)\,x^{k-1}\right)\,\frac{t^n}{n!}.
\end{align}
On the other hand, by the definition of $P_n(x)$ in (3.3) in conjunction with the recurrence relation for $P_n(x)$  (3.3*),  we have
\begin{equation} P(x,t)= P_{0}(x)+\sum_{n\,=1}^{\infty} \,P_{n}(x)\,\frac{t^n}{n!}= x + \sum_{n\,=1}^{\infty} \big(1+x^2\big)\,P_{n-1}^{'}(x)\,\frac{t^n}{n!}
\end{equation}
\noindent and thus comparing (3.7) with  (3.8) clearly yields
\begin{equation}P_{n-1}^{'}(x)= \sum_{k=1}^{n}
T(n,k)\,x^{k-1},\quad \textup{thus}\quad P_{n}(x)= P_{n}(0)+ \sum_{k=1}^{n+1}
\frac{1}{k}\,T(n+1,k)\,x^{k}.
\end{equation}

Finally, the desired result (3.4) follows from (3.9) since, by the definitions (3.2) and (3.3), we have
$P_n(0)= T(n,1)$.

\section{Concluding remarks}

It is interesting to note that several special functions, such as the  Riemann zeta function $\zeta(s),$  the Hurwitz zeta function $\zeta(s,a)$ and the classical gamma function $\Gamma(z),$  when the argument of a function takes on non-positive integer values show an abrupt changes in properties which cannot be easily explained. For instance, at $z = 0, -1, -2, \ldots,$  the function $\Gamma(z)$  becomes singular, while $\zeta(s)$ takes on rational values. On the other hand, $\zeta(s,a),$ for $s = 0, -1, -2, \ldots,$ reduces to a polynomial in $a$ which is expressible in terms of the Bernoulli polynomial $B_n(x)$ as follows
$$\zeta(- n ,a)= - \frac{B_{n+1} (a)}{n + 1}\qquad (n\in\mathbb{N}_{0}).$$
\noindent Similarly, the polypseudologs considered in this paper, {\em i.e.} the polylogs $\textup{Li}_{s} (z)$ for $s = 0, -1, -2, \ldots,$ are, in fact, rational functions in $z$.

Lee \cite{Lee03} already made use of the nil-log $\textup{Li}_{0} (z) = z/(1+z)$ in his study of the statistical mechanics of ideal particles in null dimension and discovered several anomalous physical phenomena, among them probably  the most remarkable one is that the bosons are not confinable. It remains to be seen whether it would be possible, as it is expected, to employ the polypseudologs $\textup{Li}_{-1} (z),$ $\textup{Li}_{-2} (z),$ $\textup{Li}_{-3} (z),$ $\ldots,$ in some kind of extension of Lee's formalism to the case of negative physical dimensions. It is hoped that this paper will contribute to a better understanding of the analytical properties of these functions and stimulate further work in the field.

\vskip 20mm

\section*{Acknowledgements}\noindent
{\small The author is very grateful to the three anonymous referees  for a careful and thorough reading of the previous version of this paper. Their helpful and valuable comments and suggestions have led to a considerably  improved presentation of the results. The author acknowledges financial support from Ministry of Science of the Republic of Serbia under Research Projects 142025 and 144004.}

\end{document}